\documentclass[12pt]{article}
\usepackage{amsmath,amssymb,epsf}
\usepackage{cite}
\usepackage{graphicx}
\newcommand{\beq}{\begin{equation}}
\newcommand{\eeq}{\end{equation}}
\newcommand{\bea}{\begin{eqnarray}}
\newcommand{\eea}{\end{eqnarray}}
\newcommand{\bdm}{\begin{displaymath}}
\newcommand{\edm}{\end{displaymath}}

\newcommand{\noi}{\noindent}
\newcommand{\ra}{\rightarrow}
\renewcommand{\theequation}{\thesection.\arabic{equation}}
\def\<{\langle}
\def\>{\rangle}

\def\a{\alpha}
\def\b{\beta}

\def\d{\delta}
\def\e{\epsilon}           % Also, \varepsilon
\def\g{\gamma}

\def\lam{\lambda}
\def\m{\mu}
\def\n{\nu}
  
\def\p{\pi}                % Also, \varpi
\def\th{\theta}                   %     \vartheta
\def\s{\sigma}                                   %     \varsigma

\def\D{\Delta}

\def\O{\Omega}

\def \R {{\mathbb R}}

\def\N{{\mathcal{N}}}

\begin{document}
\baselineskip=15.5pt
\pagestyle{plain}
\setcounter{page}{1}
%--------+---------+---------+---------+---------+---------+---------+

\begin{flushright}
%{\tt hep-th/yymmnnn}
\end{flushright}

\vskip 2cm

\begin{center}
{\Large \bf Hall Conductivity of Flavor Fields from AdS/CFT}
\vskip 1cm

{\bf Andy O'Bannon} \\
\vskip 0.5cm
{\it  Department of Physics, University of Washington, \\
Seattle, WA 98195-1560 \\}
{\tt  E-mail: ahob@u.washington.edu} \\
\medskip

\end{center}

\vskip1cm

\begin{center}
{\bf Abstract}
\end{center}
\medskip
We use the AdS/CFT correspondence to compute a conductivity associated with massive $\N=2$ supersymmetric hypermultiplet fields at finite baryon density, propagating through an $\N=4$ supersymmetric $SU(N_c)$ Yang-Mills plasma in the large $N_c$, large 't Hooft coupling limit. We do so by introducing external electric and magnetic fields coupled to baryon number and computing the resulting induced current, from which we extract the conductivity tensor. At large hypermultiplet mass we compute the drag force on the charge carriers. We also compute the product of the drag coefficient with the kinetic mass, and find that the answer is unchanged from the zero density case. The gravitational dual is a probe D7-brane, with a nontrivial worldvolume gauge field configuration, in an AdS-Schwarzschild background. We identify an effective horizon on the D7-brane worldvolume analogous to the worldsheet horizon observed for strings moving in the same background. We generalize our results to a class of theories described by probe D-branes in various backgrounds.

\newpage

%%%%%%%%%%%%%%%%%%%%%%%%%%%%%%%%%%%%%%%%%%%%%%%%%%%%%%%%%%%%%%%%%%%%%%%
%%%%%%%%%%%%%%%%%%%%%%%%%%%%%%%%%%%%%%%%%%%%%%%%%%%%%%%%%%%%%%%%%%%%%%%
\section{Introduction} \label{intro}
%%%%%%%%%%%%%%%%%%%%%%%%%%%%%%%%%%%%%%%%%%%%%%%%%%%%%%%%%%%%%%%%%%%%%%%
%%%%%%%%%%%%%%%%%%%%%%%%%%%%%%%%%%%%%%%%%%%%%%%%%%%%%%%%%%%%%%%%%%%%%%%

The conductivity tensor $\s_{ij}$ measures the response of a conducting medium to externally applied fields. It is defined by

\beq
\< J_i \> = \s_{ij} E_j \nonumber
\eeq

\noi where $E$ are externally applied electric fields and $\< J \>$ are the currents induced in the medium. An external magnetic field $B$ produces off-diagonal elements in $\s_{ij}$: the induced current is perpendicular to both $E$ and $B$. This is the Hall effect. For a rotationally-invariant system with $E$ in the $x$ direction and $B$ perpendicular to the $xy$ plane, $\s_{xx} = \s_{yy}$ and $\s_{xy} = -\s_{yx}$. The component $\s_{xx}$ is called the Ohmic conductivity and $\s_{xy}$ the Hall conductivity.

Our goal in this paper is to compute a conductivity associated with massive $\N=2$ supersymmetric hypermultiplet flavor fields propagating in an $\N=4$ supersymmetric $SU(N_c)$ Yang-Mills (SYM) plasma at temperature $T$. We work in the limits $N_c \ra \infty$ and 't Hooft coupling $\lam \equiv g_{YM}^2 N_c \gg 1$. We take the number $N_f$ of flavor fields to be $N_f \ll N_c$, so that for massless hypermultiplets the theory is conformal to leading order in $N_f / N_c$.

The flavor fields have a global $U(N_f)$ symmetry whose $U(1)_B$ subgroup we identify as baryon number. We work at finite $U(1)_B$ density. If we introduce non-dynamical $E$ and $B$ fields that couple to $U(1)_B$ charge, then the flavor degrees of freedom will be accelerated. The $\N=4$ SYM plasma provides resistance, allowing for a steady-state $U(1)_B$ current $J^{\mu}$. This is the origin of the conductivity we will compute. We extend the result of ref. \cite{Karch:2007pd}, where only $E$ was included, to nonzero $B$ and hence nonzero $\s_{xy}$.

Our main tool will be the anti- de Sitter / Conformal Field Theory correspondence (AdS/CFT), which equates the $\N=4$ SYM theory in the limits described above with supergravity on the ten-dimensional spacetime $AdS_5 \times S^5$ \cite{Maldacena:1997re,Witten:1998qj,Gubser:1998bc}. The SYM theory in thermal equilibrium is dual to supergravity on an AdS-Schwarzschild spacetime, where the SYM theory temperature is identified with the Hawking temperature of the AdS-Schwarzschild black hole \cite{Gubser:1996de,Witten:1998zw}. This conjectured correspondence originated from analysis of the black  D3-brane solution in type IIB string theory \cite{Maldacena:1997re}.

The $N_f$ $\N=2$ hypermultiplet fields appear in the supergravity description as $N_f$ D7-branes \cite{Karch:2002sh}. When we introduce only $N_f \ll N_c$ of them, we may neglect their back-reaction on the geometry: they are probes. The D7-brane action is then the Dirac-Born-Infeld (DBI) action. The hypermultiplet mass $m$ is dual to the geometry of the D7-brane in a way we will make precise in the sequel. The global $U(1)_B$ symmetry is dual to the $U(1)$ worldvolume gauge invariance of the D7-branes.

More specifically, if we wish to study the field theory with finite baryon number density $\< J^t \>$, we introduce a nontrivial time component $A_t(z)$ of the D7-brane gauge field, with $z$ the AdS radial coordinate \cite{Kobayashi:2006sb}. Following the usual AdS/CFT prescription \cite{Witten:1998qj,Gubser:1998bc}, $A_t(z)$'s behavior near the AdS boundary gives the $U(1)_B$ chemical potential, $\m_B$, and density, $\< J^t\>$, of the SYM theory. In the field theory we also want background electric and magnetic fields $F^{tx} = E$ and $F^{xy} = B$ and induced currents $\< J^x \>$ and $\< J^y \>$. We introduce these in the supergravity theory as nontrivial gauge field components $A_x(z,t) = -E t + f_x(z)$, which produces $E$ and $\< J^x \>$, and $A_y(z,x) = B x + f_y(z)$, which produces $B$ and $\< J^y \>$.

As shown in ref. \cite{Kobayashi:2006sb}, when $A_t(z)$ is nontrivial the only physically allowed D7-brane embeddings are the so-called ``black hole'' embeddings. These are D7-branes extended in the $AdS_5 \times S^3$ directions and intersecting the AdS-Schwarzschild horizon. They thus possess a worldvolume horizon themselves. As we want a nontrivial $A_t(z)$, we will work only with black hole D7-brane embeddings. We review D7-brane embeddings in more detail below.

To illuminate salient features of our system we will compare to refs. \cite{Hartnoll:2007ai,Hartnoll:2007ih,Herzog:2007ij,Hartnoll:2007ip}, where the conductivity tensor of a strongly-coupled, finite-temperature CFT in 2+1 dimensions was computed using gauge-gravity duality. An example of such a theory is the $\N=8$ SYM theory in 2+1 dimensions, with a $U(1)$ subgroup of the $SO(8)$ R-symmetry playing the role of electromagnetism. The gravitational dual of this theory is eleven-dimensional supergravity on $AdS_4 \times S^7$, consistently truncated to Einstein-Maxwell theory on $AdS_4$. Electric and magnetic fields in the field theory are described in the gravity theory by a dyonic black hole \cite{Hartnoll:2007ai}. In refs. \cite{Herzog:2007ij,Hartnoll:2007ip}, the external fields were given harmonic time dependence. We include only static external fields in our setup, so we will compare to the zero-frequency result of refs. \cite{Hartnoll:2007ai,Hartnoll:2007ih,Hartnoll:2007ip}, which was, in fact, identical to the result for a Lorentz-invariant system \cite{Hartnoll:2007ai} obeying linear (Maxwell) electrodynamics.

Our SYM theory differs from that of refs. \cite{Hartnoll:2007ai,Hartnoll:2007ih,Herzog:2007ij,Hartnoll:2007ip} in two important ways. First, our theory is not a CFT. Our hypermultiplet fields have the mass $m$. Second, our system effectively has energy and momentum dissipation. The flavor fields contribute an order $N_f N_c$ term to the stress-energy tensor. When $N_f \ll N_c$, this is dwarfed by the order $N_c^2$ contribution from the $\N=4$ SYM plasma. Our moving charges may thus transfer their energy and momentum into the plasma at a constant rate, without producing any significant motion of the plasma, for at least a time of order $N_c$. This is why a time-independent, steady-state solution appears in the limit of large $N_c$ with $N_f \ll N_c$.

Additionaly, our method differs from that of refs. \cite{Hartnoll:2007ai,Hartnoll:2007ih,Herzog:2007ij,Hartnoll:2007ip}. We will not use Kubo formulas to compute the conductivity, as in refs.  \cite{Hartnoll:2007ai,Hartnoll:2007ih,Herzog:2007ij,Hartnoll:2007ip}. Kubo formulas are only valid in the regime of linear response. For flavor fields, we can capture some nonlinear effects. This is because, in the supergravity description, we use a DBI action rather than a Yang-Mills action. We calculate the conductivity simply by demanding reality of the on-shell DBI action \cite{Karch:2007pd}.

If we take a limit in which $m$ is finite but arbitrarily larger than any other scale, for example the scale $\Delta m = \frac{1}{2} \sqrt{\lambda} T$ of zero-density thermal corrections to $m$ \cite{Herzog:2006gh}, we expect the flavor excitations to behave as quasi-particles. We denote this limit $m \ra \infty$. In this limit we can compute the drag force on the charge carriers, and in particular we can compute $\m M$, where $\m$ is the drag coefficient and $M$ is the kinetic mass of the quasi-particles, distinct from the Lagrangian mass $m$ at finite temperature and density\footnote{At zero density, for $m \gg \Delta m$, we know $M = m \left(1 - \frac{\Delta m}{m} + O(\frac{\Delta m}{m})^2 \right)$ \cite{Herzog:2006gh}. In our $m \ra \infty$ limit, $M$ and $m$ are therefore indistinguishable. We will continue to use the symbol $M$, however, to remind ourselves of the distinction outside of this limit.}. $\m M$ was computed in the $m \ra \infty$ limit for $\<J^t\> = 0$ in refs. \cite{Herzog:2006gh,Gubser:2006bz,Herzog:2006se} and at finite $\<J^t \>$ in ref. \cite{Karch:2007pd}. The result in both cases was $\m M = \frac{\p}{2} \sqrt{\lambda} T^2$. We will find the same answer at finite $B$. We will argue that the result is independent of the density and external fields simply because we are working to leading order in large-$N_c$. We will draw an instructive comparison, however, between our calculation and the calculation of $\m M$ from a single moving string \cite{Herzog:2006se,Herzog:2007kh}. In particular, we identify an effective horizon on the D7-brane worldvolume analogous to the worldsheet horizon on a single string \cite{Gubser:2006nz,CasalderreySolana:2007qw}.

Everything we will do comes with a caveat: the phase diagram in the parameter space of $T$, $\< J^t \>$, $E$ and $B$ (in units of $m$) is not fully known. At $E=B=0$, a region of instability is known to exist in the plane of $\<J^t\>$ versus $T$, and for sufficiently large chemical potential the hypermultiplet scalars may undergo Bose-Einstein condensation \cite{Kobayashi:2006sb}. In such regions of parameter space our D7-brane solutions do not represent the ground state of the theory and must be discarded. Our results are valid only when D7-brane black hole embeddings are the appropriate supergravity description\footnote{At finite $B$ with $T = E = \< J^t \> = 0$ the field theory exhibits spontaneous breaking of a chiral symmetry even at $m=0$ and a Zeeman-like splitting in the meson spectrum \cite{Filev:2007gb,Filev:2007qu}.}.

As in ref. \cite{Karch:2007pd}, we may also generalize our results to theories whose gravitational duals are probe Dq-branes in backgrounds of Dp-branes. This is possible when the Dq-brane has a worldvolume horizon and the Dq-brane's dynamics is described by the DBI term alone. Wess-Zumino couplings will, in general, introduce new terms into the equation of motion for the Dq-brane worldvolume gauge field that may render our solution inapplicable. 

This paper is organized as follows. In section \ref{prelims} we briefly review some results from classical electromagnetism. In section \ref{d7} we solve for the probe D7-brane gauge field. In section \ref{sigma} we compute the conductivity. In section \ref{drag} we compute $\m M$ in the $m \ra \infty$ limit. In section \ref{dpdq} we generalize our results to Dq-brane probes in Dp-brane backgrounds. We conclude in section \ref{conclusion}. In the Appendix we use holographic renormalization to compute $\< J^{\mu} \>$.

%%%%%%%%%%%%%%%%%%%%%%%%%%%%%%%%%%%%%%%%%%%%%%%%%%%%%%%%%%%%%%%%%%%%%%%
%%%%%%%%%%%%%%%%%%%%%%%%%%%%%%%%%%%%%%%%%%%%%%%%%%%%%%%%%%%%%%%%%%%%%%%
\section{Preliminaries} \label{prelims}
%%%%%%%%%%%%%%%%%%%%%%%%%%%%%%%%%%%%%%%%%%%%%%%%%%%%%%%%%%%%%%%%%%%%%%%
%%%%%%%%%%%%%%%%%%%%%%%%%%%%%%%%%%%%%%%%%%%%%%%%%%%%%%%%%%%%%%%%%%%%%%%
\setcounter{equation}{0}

We first review two results from classical electromagnetism that we will reproduce from our supergravity calculation in appropriate limits.

Imagine filling the vacuum with a charge density $\< J^t \>$. In the lab frame we may introduce a magnetic field $\vec{B}$. In a frame moving with velocity $-\vec{v}$ relative to the lab frame we will find a current $\vec{J} = \< J^t \> \vec{v}$ and an electric field

\beq
\vec{E} = - \vec{v} \times \vec{B} = - \frac{1}{\< J^t \>} \vec{J} \times \vec{B}.
\eeq

\noi If we take $\vec{B} = (0,0,B)$ we find the conductivity

\beq
\s_{xx} = 0, \qquad \s_{xy} = \< J^t \> / B.
\label{zerotemp}
\eeq

\noi Notice that this argument does not require that the charge density be comprised of quasi-particle charge carriers. Indeed, this argument relies only on Lorentz invariance. This was the result found in refs. \cite{Hartnoll:2007ai,Hartnoll:2007ih,Hartnoll:2007ip} for a (2+1)-dimensional CFT at finite temperature.

Now imagine a density $\< J^t \>$ of massive quasi-particles propagating non-relativistically through an isotropic, homogeneous, neutral medium. In the rest frame of the medium we introduce an electric field $E$ in the $\hat{x}$ direction in addition to the magnetic field. The force on a quasi-particle is then

\beq
\frac{d \vec{p}}{dt} = \vec{E} + \vec{v} \times \vec{B} - \m \vec{p},
\eeq

\noi where our quasi-particle has charge $+1$ and $\m$ is the drag coefficient. We replace the momentum with the velocity using $\vec{p} = M \vec{v}$ for quasi-particle mass $M$. We then replace the velocity with the induced current using $\vec{v} = \< \vec{J} \>/\< J^t \>$. Imposing the steady-state condition $\frac{d \vec{p}}{dt} = 0$ and solving for $\< \vec{J}\>$ yields

\beq
\s_{xx} = \frac{\s_0}{(B / \m M)^2 + 1}, \qquad \s_{xy} = \frac{\s_0 (B / \m M)}{(B / \m M)^2 +1}
\label{drude}
\eeq

\noi where $\s_0 = \< J^t \> / \m M$ is the conductivity when $B=0$.

%%%%%%%%%%%%%%%%%%%%%%%%%%%%%%%%%%%%%%%%%%%%%%%%%%%%%%%%%%%%%%%%%%%%%%%
%%%%%%%%%%%%%%%%%%%%%%%%%%%%%%%%%%%%%%%%%%%%%%%%%%%%%%%%%%%%%%%%%%%%%%%
\section{The Probe D7-Brane Solution} \label{d7}
%%%%%%%%%%%%%%%%%%%%%%%%%%%%%%%%%%%%%%%%%%%%%%%%%%%%%%%%%%%%%%%%%%%%%%%
%%%%%%%%%%%%%%%%%%%%%%%%%%%%%%%%%%%%%%%%%%%%%%%%%%%%%%%%%%%%%%%%%%%%%%%
\setcounter{equation}{0}

In type IIB string theory, we consider a system of $N_c$ non-extremal D3-branes and $N_f$ D7-branes aligned in flat ten-dimensional space as

\begin{equation}
\begin{array}{ccccccccccc}
   & X_0 & X_1 & X_2 & X_3 & X_4 & X_5 & X_6 & X_7 & X_8 & X_9\\
\mbox{D3} & \times & \times & \times & \times & & &  &  & & \\
\mbox{D7} & \times & \times & \times & \times & \times  & \times
& \times & \times &  &   \\
\end{array}
\end{equation}

\noi The $X_8$ and $X_9$ directions are orthogonal to both stacks of D-branes, which thus appear as points in the $X_8$-$X_9$ plane. If we separate these points, an open string may stretch between the two stacks. The mass of this string is its length times its tension. This mass appears in the SYM theory on the D3-brane worldvolume as the hypermultiplet mass $m$.

We take the usual AdS/CFT limit, $N_c \ra \infty$, $g_s \ra 0$ with $g_s N_c$ fixed and $g_s N_c \gg 1$ \cite{Maldacena:1997re}. We obtain the near-horizon geometry of non-extremal D3-branes, five-dimensional AdS-Schwarzschild times $S^5$. We use an AdS-Schwarzschild metric, in units where the AdS radius is one,

\beq
ds^2 = \frac{dz^2}{z^2} - \frac{1}{z^2} \frac{(1 - z^4 / z_H^4)^2}{1+z^4/ z_H^4} dt^2 + \frac{1}{z^2} (1+z^4 / z_H^4) d\vec{x}^2
\eeq

\noi where $z$ is the radial coordinate, $t$ the time coordinate and $d\vec{x}^2$ is the metric of three-dimensional Euclidean space. The boundary is at $z = 0$ and the black hole horizon is at $z = z_H$ with $z_H^{-1} = \frac{\p}{\sqrt{2}} T$. Our $S^5$ metric is

\beq
d\Omega_{5}^2 = d\theta^2 + \sin^{2}\theta d\psi^2 + \cos^{2}\theta d\Omega_{3}^2.
\eeq

\noi where $d\Omega_{3}^2$ is the standard $S^3$ metric and $\th$ runs from zero to $\p/2$. We have chosen coordinates such that $X_8 = \frac{1}{z} \sin \th$. In our units, string theory and SYM quantities are related by $\a'^{-2} = 4 \p g_s N_c = g_{YM}^2 N_c \equiv \lam$.

In the near-horizon geometry the D7-branes extend along $AdS_5 \times S^3$ \cite{Karch:2002sh}. Nonzero separation in the $X_8$-$X_9$ plane appears in the near-horizon geometry as a D7-brane with non-trivial embedding. Specifically, the position of the worldvolume $S^3$ on the $S^5$ will be described by an embedding function $\th(z)$ \cite{Karch:2002sh}. $\th(z)$ is dual holographically to the hypermultiplet mass operator\footnote{$\th(z)$ is dual to the operator given by taking $\frac{\partial}{\partial m}$ of the SYM theory Lagrangian. This operator includes the mass operator as well as couplings to adjoint scalars. The exact operator is written in ref. \cite{Kobayashi:2006sb}. Thinking in terms of the mass operator will be sufficient for our purposes.} \cite{Karch:2002sh}. $\th(z)$'s leading asymptotic value, denoted $\th_0$ in the Appendix, is simply the separation between the D3-branes and the D7-branes, hence $m = \frac{\th_0}{2\p\a'}$.

$\th(z)$ is determined by an equation of motion derived from the D7-brane action and a boundary condition, the value of $m$. At one extreme is $m=0$, which produces $\th(z)=0$, the trivial solution to the equation of motion. In this case the D7-brane wraps the maximum-volume equatorial $S^3 \subset S^5$ for all $z$. At zero temperature, nonzero $m$ produces the so-called Minkowski embeddings, in which the worldvolume $S^3$ shrinks as we move away from $z=0$ and eventually collapses to zero volume: $\th(z') = \frac{\p}{2}$ and $\cos\th(z')=0$ at some $z'$. The D7-brane then does not extend past $z'$ in the radial direction, rather, it appears to end abruptly at $z'$ \cite{Karch:2002sh}. At the other extreme is $m = \infty$, which produces $\th(z) = \frac{\pi}{2}$ for all $z$. This effectively eliminates the D7-brane, which ends right at the boundary.

\begin{figure}
{
\centering
\begin{tabular}{ccc}
\includegraphics[width=0.3\textwidth]{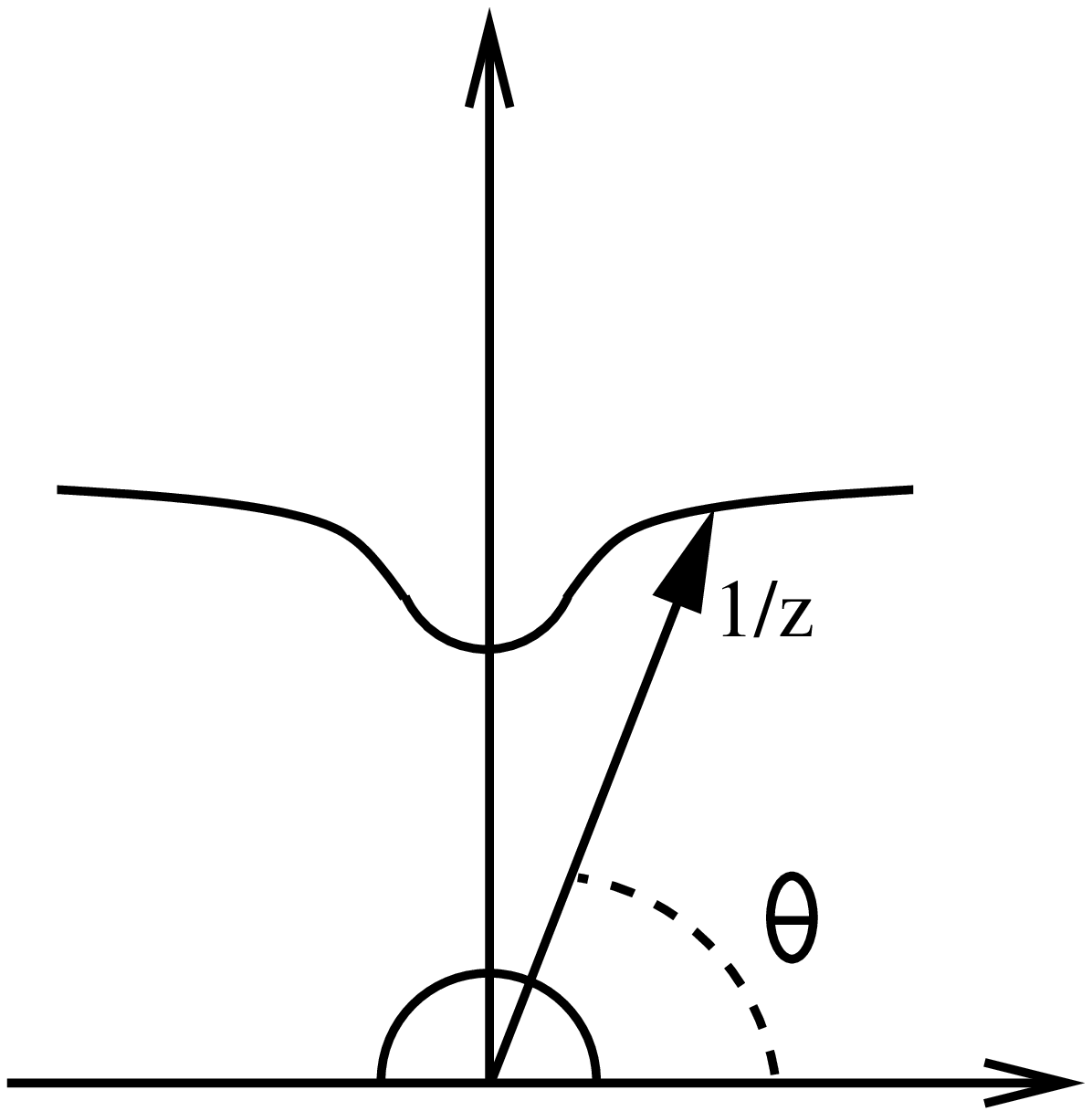} &
\includegraphics[width=0.3\textwidth]{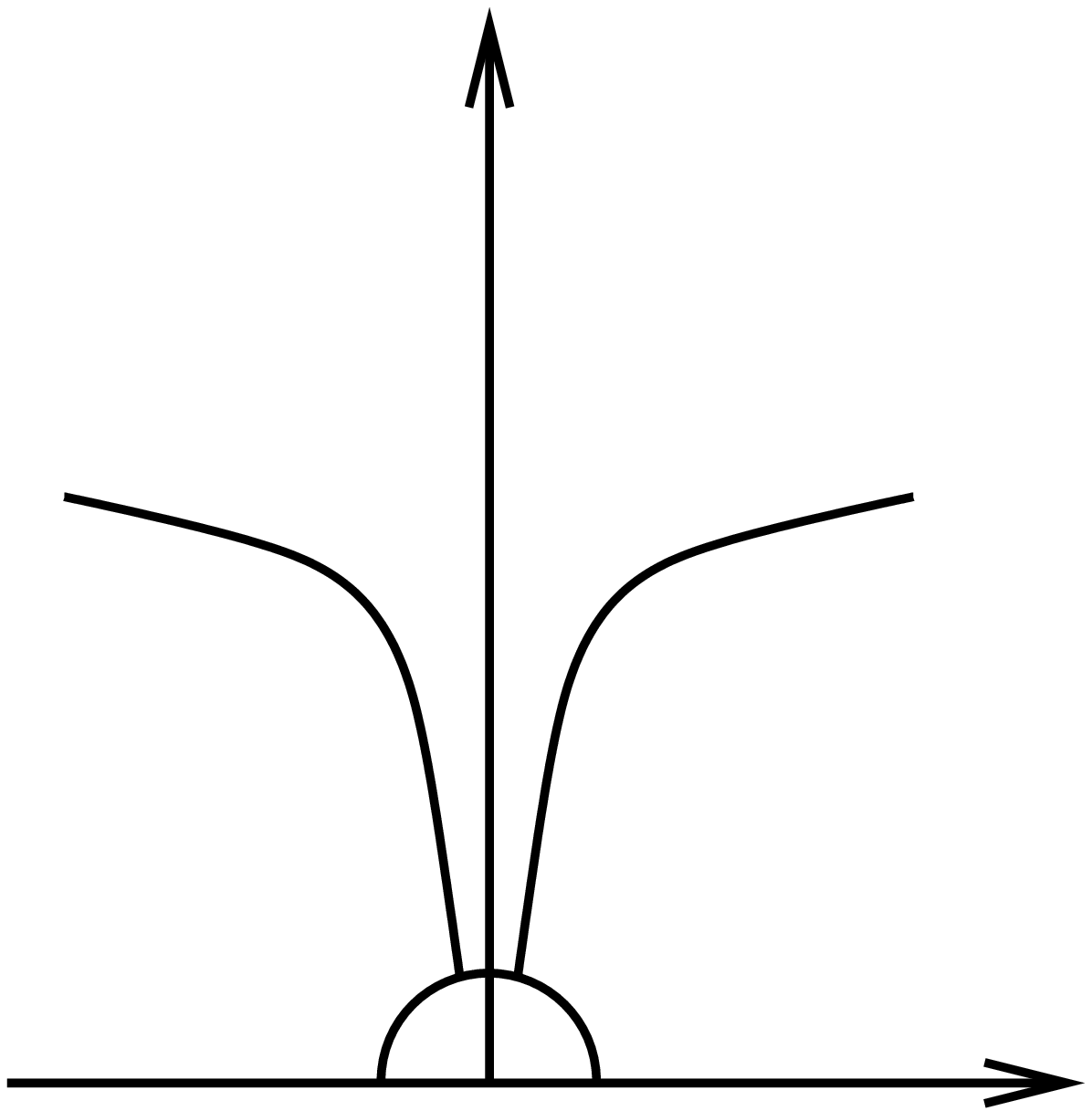}&
\includegraphics[width=0.3\textwidth]{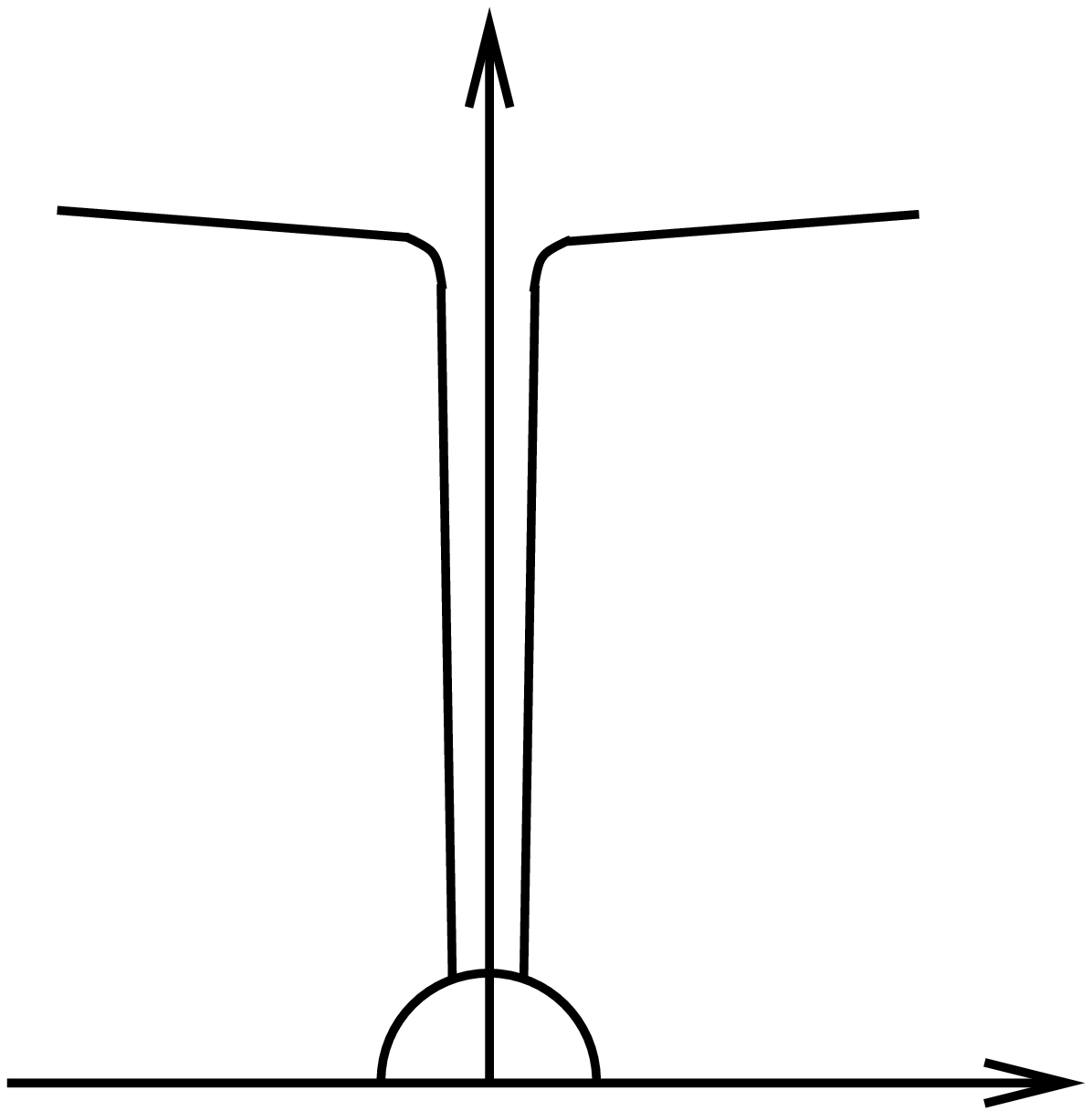}\\
(a) & (b) & (c)
\end{tabular}
\caption{\label{embeddings} Cartoons of D7-brane embeddings, with the coordinates $z$ and $\th$ indicated. We can imagine that the D3-branes sit at the origin. The semicircle about the origin represents the horizon at $z = z_H$. The boundary is $z=0$. $\th$ runs from $0$ to $\frac{\p}{2}$. The horizontal axis is a direction transverse to the D3-branes but parallel to the D7-branes, \textit{i.e.} one of $X_4, X_5, X_6$ or $X_7$. The vertical axis is $X_8$. (a.) A Minkowski embedding. (b.) A black hole embedding. (c.) A black hole embedding with a ``spike'' in the $m \ra \infty$ limit.}
}
\end{figure}

In the AdS-Schwarzschild background, with no gauge field excited on the D7-brane worldvolume, two classes of embedding are possible. The first are Minkowski embeddings that end outside the horizon, $z' < z_H$. These do not possess a horizon on their worldvolume. The second class of embeddings are black hole embeddings, in which the $S^3$ never collapses to zero volume and the D7-brane intersects the AdS-Schwarzschild horizon. The D7-brane thus possesses a horizon on its worldvolume. These embeddings are depicted in Fig. \ref{embeddings}.

If we introduce a worldvolume gauge field $A_t(z)$, the resulting radial electric field lines must have some place to end. For a Minkowski embedding no such place exists. We may introduce point sources, strings stretching from the D7-brane to the horizon, to accommodate the radial field lines. As shown in ref. \cite{Kobayashi:2006sb}, however, the force that the strings exert on the D7-brane will overcome the tension of the D7-brane, so the D7-brane will be drawn into the horizon, producing a black hole embedding. We will therefore work only with black hole embeddings, for which the field lines may end on the horizon.

With nonzero $A_t(z)$, in the SYM theory limit $m \ra \infty$, the D7-brane black hole embedding resembles a ``spike'': the $S^3$ \textit{almost} collapses to zero volume at some value $z \equiv z_{spike}$, but then remains at constant finite volume all the way to the horizon\footnote{For $E=B=0$, the position where corrections to the constant-volume solution are non-negligible is, defining $\th(z) = \frac{\p}{2} - \varepsilon$ with $\varepsilon \ll 1$ and using SYM quantities, $z_{spike}/z_H \sim \varepsilon \left( \frac{\<J^t\>}{\sqrt{\lam} N_f N_c T^3} \right)^{-1/3}$ \cite{Kobayashi:2006sb}.}. In fact, the action of the spike is identical to the action of a bundle of strings \cite{Kobayashi:2006sb}. This makes sense intuitively: a finite baryon density in the SYM theory should appear in the supergravity description as very many strings. What is perhaps surprising is that the D7-brane alone, with no strings introduced explicitly, manifests these strings itself via the spike.

As in ref. \cite{Karch:2007pd}, we will not solve for $\th(z)$ but we will consider limits. The $m=0$ limit is $\th(z)=0$. For $m \ra \infty$ we may approximate $\th(z) \approx \pi/2$ or $\cos \th(z) \approx 0$ when $z > z_{spike}$. In particular, we will use this for $z$ near the horizon.

We will now solve for the D7-brane worldvolume gauge fields. The D7-brane action is

\beq
S_{D7} = - N_f T_{D7} \int L =  - N_f T_{D7} \int d^8 \zeta \sqrt{- det \left( g_{a b} + (2 \p \a') F_{a b} \right )}
\eeq

\noi plus Wess-Zumino terms that will be zero in what we do. $T_{D7}$ is the D7-brane tension, $\zeta$ are worldvolume coordinates, $g_{a b}$ is the induced metric, and $F_{a b}$ is the $U(1)$ field strength. In our conventions, a string endpoint couples to this gauge field with coupling $+1$. In the SYM theory we also want fields $E$ and $B$, a charge density $\< J^t \>$ and induced currents $\< J^x \>$ and $\< J^y \>$. We thus introduce worldvolume gauge field components $A_t(z)$ and

\beq
A_x(z,t) = - E t + f_x(z), \qquad A_y(z,x) = B x + f_y(z)
\eeq

\noi so that at the boundary we have electric and magnetic fields $F^{tx} = E$ and $F^{xy} = B$. As part of our gauge choice we take $A_z = 0$. As our gauge fields only depend on $(z,t,x,y)$, the D7-brane action is simply a (3+1)-dimensional Born-Infeld action, with some ``extra'' factors in front from the $S^3$ and the extra spatial direction,

\beq
S_{D7} = - \N \int d^4x \cos^3\th g_{xx}^{1/2} \sqrt{- g - (2 \p \a')^2 \frac{1}{2} g F^2 - (2 \p \a')^4 \frac{1}{4} \left(  F \wedge F \right)^2}
\label{originaldbi}
\eeq

\noi The overall prefactor is, using $T_{D7} = \frac{\a'^{-4} g_s^{-1}}{(2\p)^7} = \frac{\lam N_c}{2^5 \p^6}$,

\beq
\N \equiv N_f T_{D7} 2 \p^2 = \frac{\lam}{(2 \pi)^4} N_f N_c.
\eeq

\noi with $2 \p^2$ the volume of a unit $S^3$. We have divided both sides of eq. (\ref{originaldbi}) by the volume of $\R$, defined $d^4x = dzdtdxdy$, and defined $g=g_{zz} g_{tt} g_{xx}^2$ as the determinant of the induced metric in the $(z,t,x,y)$ subspace, with $g_{zz} = 1/z^2 + \th'(z)^2$. Writing $F^2 = F^{\m\n} F_{\m\n}$, where Greek indices run over $(z,t,x,y)$, and $\tilde{F}^{\m\n} = \frac{1}{2} \e^{\m\n\a\b} F_{\a \b}$ for totally antisymmetric $\e^{\m\n\a\b}$ with $\e^{ztxy} = +1$, we have explicitly

\begin{subequations}
\label{gaugeadef}
\beq
\frac{1}{2} g F^2 = g_{xx}^2 A_t'^2 + g_{tt} g_{xx} A_x'^2 + g_{tt} g_{xx} A_y'^2 + g_{zz} g_{xx} \dot{A}_x^2 + g_{zz} g_{tt} \bar{A}_y^2
\label{a2def}
\eeq

\beq
\frac{1}{4} \left( F \wedge F\right)^2 = \left( \frac{1}{4} \tilde{F}^{\m\n} F_{\m\n} \right)^2 = \bar{A}_y^2 A_t'^2 + \dot{A}_x^2 A_y'^2 + 2 \bar{A}_y A_t' \dot{A}_x A_y'.
\label{a4def}
\eeq
\end{subequations}

\noi where dots, $\dot{A}$, denote derivatives with respect to $t$, primes, $A'$, denote derivatives with respect to $z$, and bars, $\bar{A}$, denote derivatives with respect to $x$.

The action only depends on the derivatives of $A_t(z)$, $f_x(z)$ and $f_y(z)$, so we will have three conserved charges. In the Appendix we identify these as $\< J^t \>$, $\<J^x\>$ and $\<J^y \>$,

\begin{subequations}
\label{jdef}
\beq
\N (2 \p \a')^2 g_{xx}^{1/2} \cos^3 \th \frac{-g_{xx}^2 A_t' -(2 \p \a')^2 (\bar{A}_y^2 A_t' + \bar{A}_y \dot{A}_x A_y')}{ \sqrt{- g - (2 \p \a')^2 \frac{1}{2} g F^2 - (2 \p \a')^4 \frac{1}{4} \left( F \wedge F \right)^2} } = \< J^t \>
\label{jtdef}
\eeq

\beq
\N (2 \p \a')^2 g_{xx}^{1/2} \cos^3 \th \frac{|g_{tt}| g_{xx} A_x'}{\sqrt{- g - (2 \p \a')^2 \frac{1}{2} g F^2 - (2 \p \a')^4 \frac{1}{4} \left( F \wedge F \right)^2}} = \< J^x \>
\label{jxdef}
\eeq

\beq
\N (2 \p \a')^2 g_{xx}^{1/2} \cos^3 \th \frac{|g_{tt}| g_{xx} A_y' - (2\p\a')^2 (\dot{A}_x^2 A_y' + \bar{A}_y \dot{A}_x A_t')}{\sqrt{- g - (2 \p \a')^2 \frac{1}{2} g F^2 - (2 \p \a')^4 \frac{1}{4} \left( F \wedge F \right)^2}} = \< J^y \>
\label{jydef}
\eeq
\end{subequations}

\noi Notice that the density and currents are order $\N (2\p\a')^2 \propto N_f N_c$.

With a little algebra we solve for the gauge fields from eq. (\ref{jdef}),

\beq
A_t'(z) = - \frac{\sqrt{g_{zz} |g_{tt}|}}{g_{xx}} \frac{\<J^t\> \xi - B a}{\sqrt{\xi \chi - a^2}}
\label{atsol}
\eeq

\noi where we have introduced the coefficients

\begin{subequations}
\label{coeffdefs}
\bea
\xi & = & |g_{tt}| g_{xx}^2 - (2\p\a')^2 \tilde{F}^{z\m} \tilde{F}^{z}_{\m} \nonumber \\[6pt] & = & |g_{tt}| g_{xx}^2 + (2\p\a')^2 \left ( |g_{tt}| B^2 - g_{xx} E^2 \right )
\label{xidef}
\eea
\bea
\chi & = & |g_{tt}| g_{xx}^2 \left [ \N^2 (2\p\a')^4 g_{xx} \cos^6 \th \right ] - (2\p\a')^2 \< J_{\m} \> \<J^{\m}\>
\label{chidef} \nonumber \\[6pt] & = & |g_{tt}| g_{xx}^2 \left [ \N^2 (2\p\a')^4 g_{xx} \cos^6 \th \right ] + (2\p\a')^2 \left ( |g_{tt}|\<J^t\>^2 - g_{xx} \left(  \<J^x\>^2 + \<J^y\>^2 \right) \right )
\label{chidef}
\eea
\bea
a & = & -(2\p\a')^2 \tilde{F}^{z\m} \< J_{\m}\> \nonumber \\[6pt] & = & (2\p\a')^2 (|g_{tt}| \<J^t\> B + g_{xx} \<J^y\> E)
\label{adef}
\eea
\end{subequations}

\noi Notice that $\xi$ is simply $- det(g_{ab} +(2\p\a')F_{ab})$ in the $(t,x,y)$ subspace, and that $\cos \th(z)$ appears only in $\chi$. We have written $\chi$ in a way that will make generalizing to Dp/Dq systems in section \ref{dpdq} more transparent. We also have

\beq
A_x'(z) = \sqrt{\frac{g_{zz}}{|g_{tt}|}} \frac{\< J^x \> \xi}{\sqrt{\xi \chi - a^2}}, \qquad A_y'(z) =  \sqrt{\frac{g_{zz}}{|g_{tt}|}} \frac{\< J^y \> \xi + E a}{\sqrt{\xi \chi - a^2}}
\label{axaysol}
\eeq

In the original action we may now replace the gauge fields with the conserved charges. The resulting effective action has only the single dynamical field $\th(z)$,

\beq
S_{D7} = - \N^2 (2\p\a')^2 \int d^4x \cos^6\th g_{xx}^2 \sqrt{g_{zz} |g_{tt}|} \frac{\xi}{\sqrt{\xi \chi - a^2}}
\label{effaction}
\eeq

\noi We may obtain the equation of motion for $\th(z)$ in two ways. We may derive it from the original action eq. (\ref{originaldbi}) and then plug in our gauge field solutions eqs. (\ref{atsol}) and (\ref{axaysol}), or we may Legendre transform to eliminate the gauge fields at the level of the action. The Legendre-transformed action $\hat{S}_{D7}$ is

\bea
\label{lteffaction}
\hat{S}_{D7} & = & S_{D7} - \int d^4x \left ( F_{zt} \frac{\d S_{D7}}{\d F_{zt}} + F_{zx} \frac{\d S_{D7}}{\d F_{zx}} + F_{zy} \frac{\d S_{D7}}{\d F_{zy}} \right ) \\ & = & - \frac{1}{(2\p\a')^2} \int d^4x g_{zz}^{1/2} |g_{tt}|^{-1/2} g_{xx}^{-1} \sqrt{\xi \chi - a^2} \nonumber
\eea

\noi where $\frac{\d \hat{S}_{D7}}{\d \<J^t \>} = A_t'(z)$, $\frac{\d \hat{S}_{D7}}{\d \< J^x \>} =  A_x'(z)$ and $\frac{\d \hat{S}_{D7}}{\d \< J^y \>} = A_y'(z)$ reproduce eqs. (\ref{atsol}) and (\ref{axaysol}).

Specifying the boundary conditions will then determine the D7-brane solution completely. First notice that, at the horizon, the gauge field must obey $A_t(z_H) = 0$ to be well-defined as a one-form. We are free to choose the leading asymptotic values of the fields near the boundary $z \ra 0$. We first choose the asymptotic value $\th_0$ of $\th(z)$.  The gauge fields asymptotically approach the boundary as

\begin{subequations}
\label{asymptotic}
\beq
A_t(z) = \m_B - \frac{1}{2} \frac{\< J^t \>}{\N (2\p \a')^2} z^2 + O(z^4)
\eeq

\beq
A_x(z) = -E t + c_x + \frac{1}{2} \frac{\< J^x \>}{\N (2 \p \a')^2} z^2 + O(z^4)
\eeq

\beq
A_y(z) = B x + c_y + \frac{1}{2} \frac{\< J^y \>}{\N (2 \p \a')^2} z^2 + O(z^4)
\eeq
\end{subequations}

\noi where $\m_B$, $c_x$ and $c_y$ are constants of integration. The leading asymptotic value $\m_B$ is the $U(1)_B$ chemical potential. For $A_x$ and $A_y$ we impose the boundary condition $c_x = c_y = 0$.

%%%%%%%%%%%%%%%%%%%%%%%%%%%%%%%%%%%%%%%%%%%%%%%%%%%%%%%%%%%%%%%%%%%%%%%
%%%%%%%%%%%%%%%%%%%%%%%%%%%%%%%%%%%%%%%%%%%%%%%%%%%%%%%%%%%%%%%%%%%%%%%
\section{The Conductivity} \label{sigma}
%%%%%%%%%%%%%%%%%%%%%%%%%%%%%%%%%%%%%%%%%%%%%%%%%%%%%%%%%%%%%%%%%%%%%%%
%%%%%%%%%%%%%%%%%%%%%%%%%%%%%%%%%%%%%%%%%%%%%%%%%%%%%%%%%%%%%%%%%%%%%%%
\setcounter{equation}{0}

We focus now on the quantity $\sqrt{\xi \chi - a^2}$ appearing in the effective action eq. (\ref{effaction}). As in ref. \cite{Karch:2007pd}, we will find that demanding reality of the effective action allows us to solve for $\langle J^x \rangle$ and $\langle J^y \rangle$, and hence the conductivity, in terms of $E$, $B$ and $\< J^t\>$.

In eq. (\ref{xidef}) we see that, as a function of $z$, $\xi$ has a zero: $\xi < 0$ at the horizon where $|g_{tt}|=0$, whereas $\xi >0$ near the boundary $z\ra0$. We denote the zero of $\xi$ as $z_*$,

\bea
\label{zstardef}
\frac{z_*^4}{z_H^4} & = & e^2 - b^2 + \sqrt{(e^2 - b^2)^2 + 2 (e^2 + b^2) + 1}\\ & & - \sqrt{ \left ( (e^2 - b^2) + \sqrt{(e^2 - b^2)^2 + 2 (e^2 + b^2) + 1} \right )^2 - 1}  \nonumber
\eea

\noi where we have defined the dimensionless quantities

\beq
e = \frac{1}{2} (2\p\a') E z_H^2 = \frac{E}{\frac{\p}{2}\sqrt{\lam}T^2}, \qquad b = \frac{1}{2}(2\p\a')B z_H^2 = \frac{B}{\frac{\p}{2} \sqrt{\lam}T^2}
\eeq

\noi and converted to field theory quantities. Knowing that $\xi$ is the $(t,x,y)$ part of $- det(g_{ab} +(2\p\a')F_{ab})$, we will interpret $z_*$ as an effective horizon on the D7-brane worldvolume. Notice that $z_* = z_H$ when $E=0$. We will also need $g_{xx}^2(z_*) = \p^4 T^4 {\cal F}(e,b)$ where

\beq
{\cal F}(e,b) = \frac{1}{2} \left ( 1+ e^2 - b^2 + \sqrt{(e^2 - b^2)^2 + 2 (e^2 + b^2) + 1}\right )
\eeq

\noi For later use notice that ${\cal F}(e,0) = e^2 + 1$ and ${\cal F}(0,b) = 1$.

In fact all three functions, $\xi$, $\chi$ and $a$ must share the same zero $z_*$. From eq. (\ref{chidef}) we see that at the horizon $\chi < 0$ while at the boundary $\chi > 0$, so $\chi$ also has a zero. In particular $\xi \chi > 0$ at the horizon and at the boundary. If $\xi$ and $\chi$ have distinct zeroes, then in the region between those zeroes one would change sign while the other would not, hence in that region $\xi \chi < 0$ and the effective action would be imaginary. The only consistent possibility is for $\xi$ and $\chi$ to share the zero at $z_*$. We must also have $a^2 < \xi \chi \ra 0$ as $z \ra z_*$, so that $a \ra 0$ at $z_*$ as well.

We thus set all of eqs. (\ref{coeffdefs}) to zero at $z_*$ and solve for $\< J^x \>$ and $\< J^y \>$,

\begin{subequations}
\label{cxcy}
\beq
\label{cx}
\< J^x \> = \frac{E g_{xx}}{g_{xx}^2 + (2\p\a')^2 B^2} \sqrt{(g_{xx}^2 + (2\p\a')^2 B^2) \N^2 (2\p\a')^4 g_{xx} \cos^6 \th(z_*) + (2\p\a')^2 \<J^t\>^2}
\eeq
\beq
\label{cy}
\< J^y \> = -\frac{(2\p\a')^2 \<J^t\>B}{g_{xx}^2 + (2\p\a')^2 B^2} E
\eeq
\end{subequations}

\noi with all functions of $z$ evaluated at $z_*$. Converting to field theory quantities, we find

\begin{subequations}
\label{d7sigma}
\beq
\s_{xx} = \sqrt{\frac{N_f^2 N_c^2 T^2}{16 \p^2} \frac{{\cal F}^{3/2}}{b^2 + {\cal F}}\cos^6 \th(z_*) + \frac{\rho^2 {\cal F}}{(b^2 + {\cal F})^2}}
\eeq
\beq
\s_{xy} = \frac{\rho b}{b^2 + {\cal F}}
\eeq
\end{subequations}

\noi where we have defined $\rho$ similarly to $e$ and $b$,

\beq
\rho = \frac{\< J^t \>}{\frac{\p}{2} \sqrt{\lam}T^2}
\eeq

\noi but while $e$ and $b$ are dimensionless, $\rho$ has dimension one.

As in ref. \cite{Karch:2007pd}, we interpret our result as follows. Two types of charge carriers contribute to the conductivity. The first are the charge carriers we have introduced explicitly in $\rho$. Taking $\rho=0$ leaves a nonzero $\s_{xx}$, however, so we must have another source of charge carriers. We will guess that these come from pair production in the plasma. Such pair production should depend on $m$ via a Boltzmann factor $e^{-m/T}$. The mass $m$, or equivalently $\th_0$, appears implicitly in eq. (\ref{d7sigma}) in $\cos \th(z_*)$, which should thus behave as $e^{-m/T}$. Notice $\cos \th(z_*)$ has the correct limiting behavior: $\cos \th(z_*) \ra 0$ as $m \ra \infty$, and $\cos \th(z_*) = 1$ for $m=0$. We are currently investigating whether $\cos \th(z_*)$ produces the Boltzmann factor \cite{Karch}.

We will check our answer in three limits. The first is simply to take $b \ra 0$ where ${\cal F}(e,0) = e^2 + 1$ and we immediately recover the result of ref. \cite{Karch:2007pd}.

To recover eq. (\ref{zerotemp}), we linearize in the electric field. In practical terms this means setting $e=0$, and hence ${\cal F}(0,b) = 1$, in eq. (\ref{d7sigma}). We also restore Lorentz invariance by taking $T \ra 0$. We find $\s_{xx} = 0$ and $\s_{xy} = \< J^t \>/B$, as expected.

To recover eq. (\ref{drude}), we return to finite $T$ and again linearize in the electric field. We additionally take the $m \ra \infty$ limit $\cos \th(z_*) \approx 0$. The conductivity becomes

\beq
\s_{xx} = \frac{\rho}{b^2 + 1}, \qquad \s_{xy} = \frac{\rho b}{b^2 + 1}.
\label{linearlargemass}
\eeq

\noi As shown in section \ref{drag}, in the $m \ra \infty$ limit we identify $\frac{\p}{2} \sqrt{\lam} T^2 = \m M$. We thus have $\rho = \frac{\< J^t \>}{\m M}$ and $b = \frac{B}{\m M}$, and the conductivity indeed has the form expected for quasi-particles propagating through an isotropic, homogeneous medium, eq. (\ref{drude}).

%%%%%%%%%%%%%%%%%%%%%%%%%%%%%%%%%%%%%%%%%%%%%%%%%%%%%%%%%%%%%%%%%%%%%%%
%%%%%%%%%%%%%%%%%%%%%%%%%%%%%%%%%%%%%%%%%%%%%%%%%%%%%%%%%%%%%%%%%%%%%%%
\section{The Drag Force} \label{drag}
%%%%%%%%%%%%%%%%%%%%%%%%%%%%%%%%%%%%%%%%%%%%%%%%%%%%%%%%%%%%%%%%%%%%%%%
%%%%%%%%%%%%%%%%%%%%%%%%%%%%%%%%%%%%%%%%%%%%%%%%%%%%%%%%%%%%%%%%%%%%%%%
\setcounter{equation}{0}

In the $m \ra \infty$ limit where $\cos \th \approx 0$, we expect the flavor excitations to be well-described as a collection of quasi-particles, with equation of motion

\beq
\frac{d \vec{p}}{dt} =  \vec{E} + \vec{v} \times \vec{B} - \m \vec{p},
\eeq

\noi with $v$ is the quasi-particle velocity and $\m$ the drag coefficient. Our first goal is to compute the magnitude of the drag force, $\m |\vec{p}|$. In the steady-state, $\frac{dp}{dt} = 0$. We then have

\beq
\m |\vec{p}| = \sqrt{E^2 + v^2 B^2 + 2 \vec{E}\cdot (\vec{v} \times \vec{B}) }
\eeq

\noi As $m\ra\infty$, we expect pair creation to be suppressed, so only the charge carriers in $\langle J^t \rangle$ should contribute to $\langle \vec{J} \rangle$, hence $\langle \vec{J} \rangle = \langle J^t \rangle \vec{v}$. We immediately read off $v^2 = |g_{tt}|/g_{xx}$ by setting $\chi$ to zero at $z_*$ and dropping the $\cos \th(z_*)$ term. Setting $\xi = 0$ at $z_*$ gives us

\beq
E^2 = \frac{1}{(2\p\a')^2} |g_{tt}|g_{xx} + \frac{|g_{tt}|}{g_{xx}} B^2 = \frac{1}{(2\p\a')^2} g_{xx}^2 v^2 + v^2 B^2,
\eeq

\noi Setting $a=0$ at $z_*$ gives us the component of $\vec{v}$ in the $\hat{y}$ direction,

\beq
v_y = \frac{\<J^y \>}{\< J^t\>} = - \frac{|g_{tt}|}{g_{xx}} \frac{B}{E} = - v^2 \frac{B}{E}.
\eeq

\noi We then have $2\vec{E}\cdot (\vec{v} \times \vec{B}) = 2 E B v_y = - 2 B^2 v^2$. The drag force is then

\beq
\m |\vec{p}| = \frac{1}{2\p\a'} g_{xx}(z_*) v
\label{d7force}
\eeq

\noi We can now compute $\m M$. To compare to refs. \cite{Herzog:2006gh,Gubser:2006bz,Herzog:2006se}, we employ the relativistic relation $|\vec{p}| = \gamma M v$ with $\gamma = \frac{1}{\sqrt{1-v^2}}$, and find

\beq
\m M = \frac{1}{2\p\a'} \sqrt{g_{xx}(z_*)^2-|g_{tt}(z_*)|g_{xx}(z_*)}
\label{mum}
\eeq

\noi which evaluates to $\frac{1}{\p\a'}z_H^{-2} = \frac{\p}{2} \sqrt{\lam} T^2$. This is identical to the zero density result of refs. \cite{Herzog:2006gh,Gubser:2006bz} and finite density result of ref. \cite{Karch:2007pd}, but now with nonzero $B$.

The $\<J^t\>$ independence is easy to understand\footnote{We thank L. Yaffe for the following argument.}. The plasma contains order $N_c^2$ adjoint degrees of freedom and order $N_f N_c$ flavor degrees of freedom. The flavor excitations are thus dilute in the large-$N_c$ limit. In a perturbative analysis, the flavor excitations will be more likely to scatter off of adjoint degrees of freedom than other flavor excitations. Scatterings with adjoint degrees of freedom will thus be the flavor excitations' primary mechanism for the microscopic energy loss that results in the macroscopic drag force. Introducing a density $\<J^t\>$ of order $N_f N_c$ will not change this to leading order in large-$N_c$. Increasing the stength of the coupling muddies the picture of isolated scatterings but does not affect the argument, which relies only on large-$N_c$ counting. Taking $m \ra \infty$, and in particular $m \gg \m_B$, serves only to dilute the charge carriers further. We therefore expect to recover the zero-density result at leading order in the $N_f \ll N_c$ limit.

The $B$ independence follows from this, simply because the zero-density result $\frac{\p}{2} \sqrt{\lam} T^2$ was already, curiously, independent of the quasi-particle momentum, or equivalently of $m$ and $v$ \cite{Herzog:2006gh}. As $v$ is determined by $E$ and $B$, and is the only place where $E$ and $B$ could appear in the answer, we expect the answer to be independent of $E$ and $B$.

\begin{figure}
{
\centering
\includegraphics[width=0.5\textwidth]{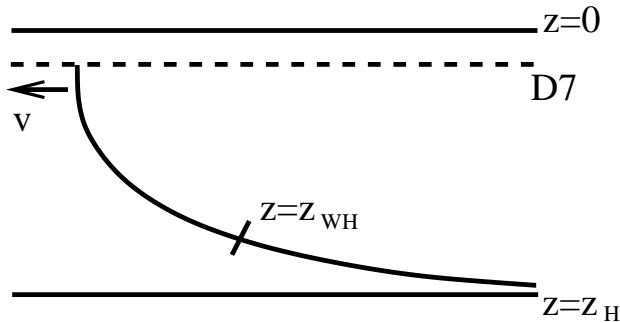}
\caption{\label{trailing} Cartoon of the trailing string. The AdS boundary $z=0$ is at the top. The AdS-Schwarzschild horizon $z=z_H$ is at the bottom. The dashed line is the position where the D7-brane ends. The worldsheet horizon on the string, $z_{WH}$, is indicated.}
}
\end{figure}

The result for the drag force, eq. (\ref{d7force}), is identical in form to the drag force computed at zero density via single-string calculations. Let us summarize the story that emerges from these single-string calculations \cite{Herzog:2006se,Herzog:2007kh,Gubser:2006nz,CasalderreySolana:2007qw}. Consider a Minkowski-embedded D7-brane that ends far from the horizon. Attach the endpoint of a string to this D7-brane. An electric field $E$ will cause this endpoint to move with velocity $v$. The body of the string will dangle into the bulk of AdS, trailing behind the endpoint (see Fig. \ref{trailing}). The string will be long and heavy, and thus behave as a classical object. Such a configuration is the single-string manifestation of our $m \ra \infty$ limit. In the SYM theory we interpret the endpoint as a single moving ``quark,'' \textit{i.e.} flavor excitation.

This ``trailing string'' in fact has a horizon on its worldsheet: a point along its length at which the time component of the induced worldvolume metric vanishes \cite{Gubser:2006nz,CasalderreySolana:2007qw,Herzog:2007kh}. Let $z_{WH}$ denote this worldsheet horizon. $z_{WH}$ is fixed by $v$. As $v \ra 1$ the horizon moves up the string, towards the boundary, while as $v \ra 0$ the horizon moves down the string, towards the AdS-Schwarzschild horizon. At $v=0$, the string stretches straight from the D7-brane to the horizon. The worldsheet horizon then coincides with the AdS-Schwarzschild horizon.

The drag force computed from such trailing strings is given by eq. (\ref{d7force}), with $z_*$ replaced by $z_{WH}$. Our effective horizon $z_*$ thus appears to be the generalization of the worldsheet horizon to the D7-brane. This makes sense intuitively when $m \ra \infty$ because the dynamics of the D7-brane spike is identical to that of a bundle of strings \cite{Kobayashi:2006sb}.

In fact, for a single string, eq. (\ref{d7force}) with $z_*$ replaced by $z_{WH}$ is the result for any asymptotically AdS geometry with a horizon \cite{Herzog:2006se}. In this sense eq. (\ref{d7force}) is ``universal,'' when written in terms of the supergravity quantity $g_{xx}$. The conversion to SYM quantities will not always reproduce $\frac{\p}{2} \sqrt{\lam} T^2$, however. For example, the charged AdS-Schwarzschild black hole background, dual to $\N=4$ SYM with nonzero R-charge density \cite{Gubser:1998jb,Chamblin:1999tk}, will produce a $\m M$ that depends on R-charge chemical potentials. That eq. (\ref{d7force}) could be ``universal'' also for the D7-brane seems plausible, but to show this would require a more general analysis. Notice, however, that the argument showing that $\xi$, $\chi$ and $a$ share a zero at $z_*$ required only that the relevant part of the D7-brane metric be asymptotically AdS and possess a horizon.

%%%%%%%%%%%%%%%%%%%%%%%%%%%%%%%%%%%%%%%%%%%%%%%%%%%%%%%%%%%%%%%%%%%%%%%
%%%%%%%%%%%%%%%%%%%%%%%%%%%%%%%%%%%%%%%%%%%%%%%%%%%%%%%%%%%%%%%%%%%%%%%
\section{Generalization to Dp/Dq Systems} \label{dpdq}
%%%%%%%%%%%%%%%%%%%%%%%%%%%%%%%%%%%%%%%%%%%%%%%%%%%%%%%%%%%%%%%%%%%%%%%
%%%%%%%%%%%%%%%%%%%%%%%%%%%%%%%%%%%%%%%%%%%%%%%%%%%%%%%%%%%%%%%%%%%%%%%
\setcounter{equation}{0}

As in ref. \cite{Karch:2007pd}, we can compute the conductivity for a class of field theories whose holographic duals are probe Dq-branes in a background of Dp-branes \cite{Arean:2006pk,Myers:2006qr}. This is possible because we required only that the DBI action be a reliable effective action and that the Dq-brane had a horizon. The dual field theories will be large-$N_c$ Yang-Mills theories with $N_f \ll N_c$ fundamental-representation fields that, in some cases, may be confined to a defect. 

The Dp-brane solution includes coordinates parallel to the Dp-branes and spherical coordinates for directions transverse to the Dp-branes. In this background we may generically write the induced Dq-brane metric as

\beq
ds_{Dq}^2 = g_{zz} dz^2 + g_{tt} dt^2 + g_{xx} d\vec{x}^2 + g_{SS} d \O_n^2
\label{dqmetric}
\eeq

\noi where $z$ is the radial coordinate. We assume this induced metric depends only on $z$ and parameters like $T$. The Dq-brane wraps some $n$-sphere $S^n$ with metric component $g_{SS}$ in the space transverse to the Dp-branes. The Dq-brane worldvolume then includes $\R^{d}$ with $d=q-n-1$. A magnetic field is only possible for $d \geq 2$. We assume the Dq-brane worldvolume has a horizon $z_H$ defined by $g_{tt}(z_H) = 0$.  The Dp-brane background may also include a nontrivial dilaton $\phi(z)$ and nontrivial Ramond-Ramond (RR) form fields.

We now introduce $A_t(z)$, $A_x(z,t)$ and $A_y(z,x)$. The Dq-brane action includes the Born-Infeld term and Wess-Zumino couplings to background RR fields. The Born-Infeld term is again a (3+1)-dimensional Born-Infeld action with an ``extra'' factor,

\beq
S_{Dq} =  - \int d^4x \frac{c(z)}{(2\p\a')^2} \sqrt{- g - (2 \p \a')^2 \frac{1}{2} g F^2 - (2 \p \a')^4 \frac{1}{4} \left(  F \wedge F \right)^2},
\label{dqdbi}
\eeq

\noi where we have divided both sides by the volume of $\R^{d-2}$, and now the ``extra'' factor is

\beq
c(z) = \N_q (2\p\a')^2 e^{-\phi(z)} g_{xx}^{\frac{d}{2}-1}(z) g_{SS}^{n/2}(z).
\eeq

\noi where $\N_q \equiv N_f T_{Dq} V_n$, with $T_{Dq}$ is the Dq-brane tension and $V_n$ is the volume of a unit $S^n$. Comparing eqs. (\ref{dqdbi}) and (\ref{originaldbi}) we see that everything is identical to what we have already done, but with

\beq
\N (2\p\a')^2 g_{xx}^{1/2} \cos^3\th \ra c(z)
\eeq

\noi In particular, the only change in eq. (\ref{coeffdefs}) is in $\chi$,

\beq
\chi = |g_{tt}| g_{xx}^2 c(z)^2 + (2\p\a')^2 \left ( |g_{tt}|\< J^t\>^2 - g_{xx}\left( \<J^x\>^2 + \<J^y \>^2 \right) \right ) 
\eeq

\noi In the Appendix we show that the identification of $\< J^t\>$, $\< J^x\>$ and $\< J^y\>$ is valid for any probe Dq-brane satisfying our assumptions, so taking $\xi = \chi = a = 0$ at $z_*$ we find

\begin{subequations}
\label{dqsigma}
\beq
\s_{xx} = \frac{g_{xx}}{g_{xx}^2 + (2\p\a')^2 B^2} \sqrt{(g_{xx}^2 + (2\p\a')^2 B^2) c(z_*)^2 + (2\p\a')^2 \<J^t\>^2}
\eeq
\beq
\s_{xy} = \frac{(2\p\a')^2 \<J^t\> B}{g_{xx}^2 + (2\p\a')^2 B^2}
\eeq
\end{subequations}

The Dq-brane action also includes Wess-Zumino couplings to RR fields. Generically, these introduce additional terms in the gauge field equation of motion. Whether our solution remains valid must be determined on a case-by-case basis. For example, in the D4/D8/\={D8} system \cite{Sakai:2004cn}, the D8-brane action includes $\int dC_3 \wedge A \wedge F \wedge F$, with $dC_3$ proportional to the volume form of $S^4$. This coupling introduces an additional term in the equation of motion that invalidates our gauge field solution.

A flat $C_{q-3}$ form, $dC_{q-3} = 0$, will leave the Dp-brane background unchanged and produce a term, $\int C_{q-3} \wedge F \wedge F$, in the Dq-brane action that leaves the gauge field equation of motion unchanged. Our solution thus remains valid. Integrating $C_{q-3}$ produces a $\theta$ parameter,

\beq
S_{Dq}^{\theta} = - \frac{\theta}{8\p^2} \int F \wedge F
\eeq

\noi For our gauge field solutions this shifts $\< J^{\mu} \> \ra \< J^{\mu} \> + \D \< J^{\mu} \>$ with

\beq
\label{tshift}
\D \<J^t\> = + \frac{\th}{4\pi^2} B, \qquad \D \< J^x \> = + \frac{\th}{4\pi^2} E_y, \qquad \D \< J^y \> = - \frac{\th}{4\pi^2} E_x.
\eeq

\noi which we implement in eq. (\ref{dqsigma}) by taking $\< J^t \> \ra \<J^t \> + \frac{\th}{4\pi^2} B$ and $\s_{xy} \ra \s_{xy} + \frac{\th}{4\p^2}$.

%%%%%%%%%%%%%%%%%%%%%%%%%%%%%%%%%%%%%%%%%%%%%%%%%%%%%%%%%%%%%%%%%%%%%%%
%%%%%%%%%%%%%%%%%%%%%%%%%%%%%%%%%%%%%%%%%%%%%%%%%%%%%%%%%%%%%%%%%%%%%%%
\section{Conclusion} \label{conclusion}
%%%%%%%%%%%%%%%%%%%%%%%%%%%%%%%%%%%%%%%%%%%%%%%%%%%%%%%%%%%%%%%%%%%%%%%
%%%%%%%%%%%%%%%%%%%%%%%%%%%%%%%%%%%%%%%%%%%%%%%%%%%%%%%%%%%%%%%%%%%%%%%

Using the AdS/CFT correspondence, we computed the Hall conductivity of a finite baryon number density of $\N=2$ hypermultiplet excitations in an $\N=4$ SYM plasma in the limits of large $N_c$ and large 't Hooft coupling. Our method is valid for any values of $m$, $\< J^t \>$, $T$, $B$ and $E$ for which the supergravity description as a probe D7-brane with worldvolume horizon is valid. We also computed the drag force on flavor excitations in the plasma in the $m\ra \infty$ limit, and identified the D7-brane analogue of the trailing string worldsheet horizon.

Electric-magnetic self-duality, or S-duality, of $U(1)$ Yang-Mills theory in $AdS_4$, and its interpretation in the dual (2+1)-dimensional CFT, was studied in refs. \cite{Witten:2003ya,Leigh:2003ez,Yee:2004ju,Petkou:2004nu,deHaro:2007eg}. Put briefly, S-duality appears in the CFT as particle-vortex duality. S-duality may be extended to $SL(2,\mathbb{Z})$ if a T transformation can be found. For abelian Yang-Mills in $AdS_4$, this arises as a $2\p$ shift of the bulk $\theta$ angle, which appears in the dual field theory as a shift in the two-point function of the dual current by a contact term \cite{Witten:2003ya}. The transformation of the conductivity (and other transport coefficients) under S- and T-duality was studied in refs. \cite{Hartnoll:2007ai,Hartnoll:2007ih,Herzog:2007ij,Hartnoll:2007ip}.

A similar analysis should be possible for probe Dq-branes using the well-known extension of S-duality to (3+1)-dimensional Born-Infeld theory \cite{Gaillard:1981rj,Gibbons:1995cv,Tseytlin:1996it,Gaillard:1997rt}. Indeed, the Dq-brane action eq. (\ref{dqdbi}) is simply the (3+1)-dimensional Born-Infeld action with the extra factor $c(z)$. The $\theta$ angle we identified in the Dq-brane action produces the T transformation in the same fashion as for Yang-Mills theory.

In the condensed matter physics literature, an $SL(2,\mathbb{Z})$ duality transformation has been proposed to relate transitions between quantum Hall plateaux. As a small sampling of this literature see refs. \cite{Lutken:1991jk,Lutken:1992xu,Burgess:2000kj,Burgess:2001sy}. We note in passing that in ref. \cite{Burgess:2001sy} the $SL(2,\mathbb{Z})$ action was shown to persist unaltered even beyond the linear response regime.

We reiterate the comment of ref. \cite{Hartnoll:2007ip}, however, that how a quantum Hall effect may occur in gauge-gravity duality is currently unclear. The fundamental problem seems to be how to describe a Fermi surface using gauge-gravity duality\footnote{For recent work in this direction, see ref. \cite{Rozali:2007rx}.}. This is perhaps the most exciting direction for future research.

%%%%%%%%%%%%%%%%%%%%%%%%%%%%%%%%%%%%%%%%%%%%%%%%%%%%%%%%%%%%%%%%%%%%%%%
%%%%%%%%%%%%%%%%%%%%%%%%%%%%%%%%%%%%%%%%%%%%%%%%%%%%%%%%%%%%%%%%%%%%%%%
\section*{Acknowledgements}
%%%%%%%%%%%%%%%%%%%%%%%%%%%%%%%%%%%%%%%%%%%%%%%%%%%%%%%%%%%%%%%%%%%%%%%
%%%%%%%%%%%%%%%%%%%%%%%%%%%%%%%%%%%%%%%%%%%%%%%%%%%%%%%%%%%%%%%%%%%%%%%

We would like to thank C. Herzog, A. Karch, D. T. Son, L. Yaffe and D. Yamada for useful conversations and for reading the manuscript. This work was supported by the Jack Kent Cooke Foundation.

%%%%%%%%%%%%%%%%%%%%%%%%%%%%%%%%%%%%%%%%%%%%%%%%%%%%%%%%%%%%%%%%%%%%%%%
%%%%%%%%%%%%%%%%%%%%%%%%%%%%%%%%%%%%%%%%%%%%%%%%%%%%%%%%%%%%%%%%%%%%%%%
\section*{Appendix: Holographic Renormalization} \label{holorg}
%%%%%%%%%%%%%%%%%%%%%%%%%%%%%%%%%%%%%%%%%%%%%%%%%%%%%%%%%%%%%%%%%%%%%%%
%%%%%%%%%%%%%%%%%%%%%%%%%%%%%%%%%%%%%%%%%%%%%%%%%%%%%%%%%%%%%%%%%%%%%%%
\setcounter{equation}{0}
\renewcommand{\theequation}{\arabic{equation}}

In AdS/CFT, we equate the on-shell supergravity action with the generating functional of field theory correlation functions. The on-shell action, however, is divergent due to the radial integration. In holographic renormalization (holo-rg) \cite{Henningson:1998gx,Henningson:1998ey,Balasubramanian:1999re,deHaro:2000xn} we introduce a regulator $z=\e$, add counterterms at $z=\e$ to cancel the divergences, and then take $\e \ra 0$.

We find from its equation of motion that $\th(z)$ has the asymptotic expansion

\beq
\th(z) = \th_0 z + \th_2 z^3 + \ldots.
\label{thetasymptotic}
\eeq

\noi The leading coefficient $\th_0$ is the source for the dual operator, given by taking $\frac{\partial}{\partial m}$ of the SYM Lagrangian. In other words $\th_0$ gives the hypermultiplet mass. If we separate the D3-branes and the D7-branes by a distance $L$ in the $X_8$ direction, then $m = \frac{L}{2\p\a'}$ and $L = \lim_{z\ra0} \frac{1}{z} \sin \th(z) = \th_0$ allows us to identify $\th_0 = (2\p\a') m$.

Plugging eq. (\ref{thetasymptotic}) into the regulated action we find the divergences

\beq
S_{reg} = - \int_{\e}^{z_H} dz L = - \N \int_{\e}^{z_H} dz \left ( z^{-5} - \th_0^2 z^{-3} + \frac{1}{2} (2 \p \a')^2 (B^2 - E^2) z^{-1} + O(z) \right )
\eeq

\noi The counterterms we need are \cite{Karch:2005ms}

\beq
L_1 = \frac{1}{4} \N \sqrt{\g}, \qquad L_2 = - \frac{1}{2} \N \sqrt{\g} \th(\e)^2, \qquad L_f = \N \frac{5}{12} \sqrt{-\g} \th(\e)^4
\eeq

\noi with $\g_{ij}$ the induced metric at $z=\e$ and $\g$ its determinant. Notice that $\sqrt{-\g} = \e^{-4} + O(\e^4)$. Supersymmetry requires the finite counterterm $L_f$ \cite{Karch:2005ms}. We suppress $\int dt dx dy$ unless stated otherwise. The last divergence requires a counterterm

\beq
L_F = - \frac{1}{4} \N (2 \p \a')^2 \sqrt{\g} F^{ij} F_{ij} \log \e  = - \frac{1}{2} \N (2 \p \a')^2 (B^2 - E^2) \log \e + O(\e^4 \log \e)
\eeq

\noi The generating functional of the field theory is the $\e \ra 0$ limit of $S = S_{reg} + \sum_i L_i$. We want the expectation values $\langle J^t \rangle$, $\langle J^x \rangle$ and $\langle J^y \rangle$. In holo-rg, $\langle J^{\m} \rangle$ is

\beq
\langle J^{\m} \rangle = \lim_{\e \ra 0} \frac{1}{\e^4} \frac{1}{\sqrt{\g}} \frac{\d S}{\d A_{\m}(\e)}
\eeq

For $\<J^t\>$, we need

\beq
\d S = - \int_{\e}^{z_H} dz \frac{\d L}{\d \partial_z A_t} \partial_z \d A_t = - \frac{\d L}{\d \partial_z A_t} \int_{\e}^{z_H} dz \partial_z \d A_t = - \frac{\d L}{\d \partial_z A_t} \left ( \d A_t(z_H) - \d A_t(\e) \right ),
\eeq

\noi where we have used the fact that $\frac{\d L}{\d \partial_z A_t}$ is $z$-independent on-shell. Enforcing $\d A_t(z_H)=0$ we find $\frac{\d S}{\d A_t(\e)}= \frac{\d L}{\d \partial_z A_t}$ and hence $\langle J^t \rangle = \frac{\d L}{\d \partial_z A_t}$.

For $\langle J^x \rangle$, we reinstate $\int dt$ because $A_x$ is time-dependent,

\beq
\d S = - \int dz dt \left ( \frac{\d L}{\d \partial_z A_x} \partial_z \d A_x + \frac{\d L}{\d \partial_t A_x} \partial_t \d A_x \right )
\eeq

\noi We employ precisely the same argument as before for the first term. For the second term we observe that $\frac{\d L}{\d \partial_t A_x}$ is $t$-independent on-shell and hence

\beq
\int dt \frac{\d L}{\d \partial_t A_x} \partial_t \d A_x = \frac{\d L}{\d \partial_t A_x} \int dt \partial_t \d A_x = 0
\eeq

\noi where we demand that the fluctuation be well-behaved (vanishing) at $t = \pm \infty$. The counterterm $L_F$ gives a vanishing contribution to $\langle J^x \rangle$ for the same reason,

\bea
\d L_F & = & - \frac{1}{4} \N (2 \p \a')^2 \sqrt{\g} \g^{ij}\g^{kl} \int dt \frac{\d }{\d \partial_t A_x} \left( F_{ik} F_{jl} \right) \partial_t \d A_x \log \e \\ & = & + \frac{1}{2} \N (2 \p \a')^2 \int dt \dot{A}_x(\e) \partial_t \d A_x \log \e + O(\e^4 \log \e) \nonumber \\ & = & O(\e^4 \log \e) \nonumber
\eea

\noi We then have $\frac{\d S}{\d A_x(\e)}= \frac{\d L}{\d \partial_z A_x}$ and hence $\langle J^x \rangle = \frac{\d L}{\d \partial_z A_x}$.

$\langle J^y \rangle$ is very similar. $A_y$ depends on $x$ so we reinstate $\int dx$. We have

\beq
\d S = - \int dz dx \left ( \frac{\d L}{\d \partial_z A_y} \partial_z \d A_y + \frac{\d L}{\d \partial_x A_y} \partial_x \d A_y \right )
\eeq

\noi The same argument as above applies for the first term, and for the second term we observe that $\frac{\d L}{\d \partial_x A_y}$ is $x$-independent on-shell. Demanding that the fluctuation be well-behaved at $x = \pm \infty$ gives $\int dx \partial_x \d A_y = 0$ and no contribution from $L_F$. We thus have $\langle J^y \rangle = \frac{\d L}{\d \partial_z A_y}$.

As in ref. \cite{Karch:2007pd}, we claim that these results are valid for any probe Dq-brane with a worldvolume horizon in a Dp-brane background. The identification of $\langle J^t \rangle$ depended only on the difference in the value of $A_t$ at the horizon and its asymptotic value. This behavior will be true for any probe brane with horizon. Similar statements apply for the identifications of $\langle J^x \rangle$ and $\langle J^y \rangle$. Additional counterterms may appear for different systems but no such counterterms can change these results. Any counterterm must be built from gauge- and Lorentz-invariant combinations of the field strength. The only components of the field strength that could contribute are $F_{tx}$ and $F_{xy}$, which in our solution are constants, so we will always end up with $\int dt \partial_t \d A_x = 0$ and $\int dx \partial_x \d A_y = 0$, as above.


\begin{thebibliography}{99}

%\cite{Karch:2007pd}
\bibitem{Karch:2007pd}
  A.~Karch and A.~O'Bannon,
  ``Metallic AdS/CFT,''
  arXiv:0705.3870 [hep-th].
  %%CITATION = ARXIV:0705.3870;%%

%\cite{Maldacena:1997re}
\bibitem{Maldacena:1997re}
  J.~M.~Maldacena,
  ``The large N limit of superconformal field theories and supergravity,''
  Adv.\ Theor.\ Math.\ Phys.\  {\bf 2}, 231 (1998)
  [Int.\ J.\ Theor.\ Phys.\  {\bf 38}, 1113 (1999)]
  [arXiv:hep-th/9711200].
  %%CITATION = IJTPB,38,1113;%%

%\cite{Witten:1998qj}
\bibitem{Witten:1998qj}
  E.~Witten,
  ``Anti-de Sitter space and holography,''
  Adv.\ Theor.\ Math.\ Phys.\  {\bf 2}, 253 (1998)
  [arXiv:hep-th/9802150].
  %%CITATION = 00203,2,253;%%

%\cite{Gubser:1998bc}
\bibitem{Gubser:1998bc}
  S.~S.~Gubser, I.~R.~Klebanov and A.~M.~Polyakov,
  ``Gauge theory correlators from non-critical string theory,''
  Phys.\ Lett.\  B {\bf 428}, 105 (1998)
  [arXiv:hep-th/9802109].
  %%CITATION = PHLTA,B428,105;%%

%\cite{Gubser:1996de}
\bibitem{Gubser:1996de}
  S.~S.~Gubser, I.~R.~Klebanov and A.~W.~Peet,
  ``Entropy and Temperature of Black 3-Branes,''
  Phys.\ Rev.\  D {\bf 54}, 3915 (1996)
  [arXiv:hep-th/9602135].
  %%CITATION = PHRVA,D54,3915;%%

%\cite{Witten:1998zw}
\bibitem{Witten:1998zw}
  E.~Witten,
  ``Anti-de Sitter space, thermal phase transition, and confinement in  gauge
  theories,''
  Adv.\ Theor.\ Math.\ Phys.\  {\bf 2}, 505 (1998)
  [arXiv:hep-th/9803131].
  %%CITATION = 00203,2,505;%%

%\cite{Karch:2002sh}
\bibitem{Karch:2002sh}
  A.~Karch and E.~Katz,
  ``Adding flavor to AdS/CFT,''
  JHEP {\bf 0206}, 043 (2002)
  [arXiv:hep-th/0205236].
  %%CITATION = JHEPA,0206,043;%%

%\cite{Kobayashi:2006sb}
\bibitem{Kobayashi:2006sb}
  S.~Kobayashi, D.~Mateos, S.~Matsuura, R.~C.~Myers and R.~M.~Thomson,
  ``Holographic phase transitions at finite baryon density,''
  arXiv:hep-th/0611099.
  %%CITATION = HEP-TH/0611099;%%

%\cite{Hartnoll:2007ai}
\bibitem{Hartnoll:2007ai}
  S.~A.~Hartnoll and P.~Kovtun,
  ``Hall conductivity from dyonic black holes,''
  arXiv:0704.1160 [hep-th].
  %%CITATION = ARXIV:0704.1160;%%

%\cite{Hartnoll:2007ih}
\bibitem{Hartnoll:2007ih}
  S.~A.~Hartnoll, P.~K.~Kovtun, M.~Muller and S.~Sachdev,
  ``Theory of the Nernst effect near quantum phase transitions in condensed
  matter, and in dyonic black holes,''
  arXiv:0706.3215 [cond-mat.str-el].
  %%CITATION = ARXIV:0706.3215;%%

%\cite{Hartnoll:2007ip}
\bibitem{Hartnoll:2007ip}
  S.~A.~Hartnoll and C.~P.~Herzog,
  ``Ohm's Law at strong coupling: S duality and the cyclotron resonance,''
  arXiv:0706.3228 [hep-th].
  %%CITATION = ARXIV:0706.3228;%%

%\cite{Herzog:2007ij}
\bibitem{Herzog:2007ij}
  C.~P.~Herzog, P.~Kovtun, S.~Sachdev and D.~T.~Son,
  ``Quantum critical transport, duality, and M-theory,''
  Phys.\ Rev.\  D {\bf 75}, 085020 (2007)
  [arXiv:hep-th/0701036].
  %%CITATION = PHRVA,D75,085020;%%

%\cite{Herzog:2006gh}
\bibitem{Herzog:2006gh}
  C.~P.~Herzog, A.~Karch, P.~Kovtun, C.~Kozcaz and L.~G.~Yaffe,
  ``Energy loss of a heavy quark moving through N = 4 supersymmetric
  Yang-Mills plasma,''
  JHEP {\bf 0607}, 013 (2006)
  [arXiv:hep-th/0605158].
  %%CITATION = JHEPA,0607,013;%%

%\cite{Gubser:2006bz}
\bibitem{Gubser:2006bz}
  S.~S.~Gubser,
  ``Drag force in AdS/CFT,''
  Phys.\ Rev.\  D {\bf 74}, 126005 (2006)
  [arXiv:hep-th/0605182].
  %%CITATION = PHRVA,D74,126005;%%

%\cite{Herzog:2006se}
\bibitem{Herzog:2006se}
  C.~P.~Herzog,
  ``Energy loss of heavy quarks from asymptotically AdS geometries,''
  JHEP {\bf 0609}, 032 (2006)
  [arXiv:hep-th/0605191].
  %%CITATION = JHEPA,0609,032;%%

%\cite{Herzog:2007kh}
\bibitem{Herzog:2007kh}
  C.~P.~Herzog and A.~Vuorinen,
  ``Spinning Dragging Strings,''
  arXiv:0708.0609 [hep-th].
  %%CITATION = ARXIV:0708.0609;%%

%\cite{Gubser:2006nz}
\bibitem{Gubser:2006nz}
  S.~S.~Gubser,
  ``Jet-quenching and momentum correlators from the gauge-string duality,''
  arXiv:hep-th/0612143.
  %%CITATION = HEP-TH/0612143;%%

%\cite{CasalderreySolana:2007qw}
\bibitem{CasalderreySolana:2007qw}
  J.~Casalderrey-Solana and D.~Teaney,
  ``Transverse momentum broadening of a fast quark in a N = 4 Yang Mills
  plasma,''
  JHEP {\bf 0704}, 039 (2007)
  [arXiv:hep-th/0701123].
  %%CITATION = JHEPA,0704,039;%%

%\cite{Filev:2007gb}
\bibitem{Filev:2007gb}
  V.~G.~Filev, C.~V.~Johnson, R.~C.~Rashkov and K.~S.~Viswanathan,
  ``Flavoured large N gauge theory in an external magnetic field,''
  arXiv:hep-th/0701001.
  %%CITATION = HEP-TH/0701001;%%

%\cite{Filev:2007qu}
\bibitem{Filev:2007qu}
  V.~G.~Filev,
  ``Criticality, Scaling and Chiral Symmetry Breaking in External Magnetic
  Field,''
  arXiv:0706.3811 [hep-th].
  %%CITATION = ARXIV:0706.3811;%%

%\cite{Karch}
\bibitem{Karch}
  A.~Karch and A.~O'Bannon,
  in progress.

%\cite{Gubser:1998jb}
\bibitem{Gubser:1998jb}
  S.~S.~Gubser,
  ``Thermodynamics of spinning D3-branes,''
  Nucl.\ Phys.\  B {\bf 551}, 667 (1999)
  [arXiv:hep-th/9810225].
  %%CITATION = NUPHA,B551,667;%%

%\cite{Chamblin:1999tk}
\bibitem{Chamblin:1999tk}
  A.~Chamblin, R.~Emparan, C.~V.~Johnson and R.~C.~Myers,
  ``Charged AdS black holes and catastrophic holography,''
  Phys.\ Rev.\  D {\bf 60}, 064018 (1999)
  [arXiv:hep-th/9902170].
  %%CITATION = PHRVA,D60,064018;%%

%\cite{Arean:2006pk}
\bibitem{Arean:2006pk}
  D.~Arean and A.~V.~Ramallo,
  ``Open string modes at brane intersections,''
  JHEP {\bf 0604}, 037 (2006)
  [arXiv:hep-th/0602174].
  %%CITATION = JHEPA,0604,037;%%

%\cite{Myers:2006qr}
\bibitem{Myers:2006qr}
  R.~C.~Myers and R.~M.~Thomson,
  ``Holographic mesons in various dimensions,''
  JHEP {\bf 0609}, 066 (2006)
  [arXiv:hep-th/0605017].
  %%CITATION = JHEPA,0609,066;%%

%\cite{Sakai:2004cn}
\bibitem{Sakai:2004cn}
  T.~Sakai and S.~Sugimoto,
  ``Low energy hadron physics in holographic QCD,''
  Prog.\ Theor.\ Phys.\  {\bf 113}, 843 (2005)
  [arXiv:hep-th/0412141].
  %%CITATION = PTPKA,113,843;%%

%\cite{Witten:2003ya}
\bibitem{Witten:2003ya}
  E.~Witten,
  ``SL(2,Z) action on three-dimensional conformal field theories with Abelian
  symmetry,''
  arXiv:hep-th/0307041.
  %%CITATION = HEP-TH/0307041;%%

%\cite{Leigh:2003ez}
\bibitem{Leigh:2003ez}
  R.~G.~Leigh and A.~C.~Petkou,
  ``SL(2,Z) action on three-dimensional CFTs and holography,''
  JHEP {\bf 0312}, 020 (2003)
  [arXiv:hep-th/0309177].
  %%CITATION = JHEPA,0312,020;%%

%\cite{Yee:2004ju}
\bibitem{Yee:2004ju}
  H.~U.~Yee,
  ``A note on AdS/CFT dual of SL(2,Z) action on 3D conformal field theories
  with U(1) symmetry,''
  Phys.\ Lett.\  B {\bf 598}, 139 (2004)
  [arXiv:hep-th/0402115].
  %%CITATION = PHLTA,B598,139;%%

%\cite{Petkou:2004nu}
\bibitem{Petkou:2004nu}
  A.~C.~Petkou,
  ``Holography, duality and higher-spin theories,''
  arXiv:hep-th/0410116.
  %%CITATION = HEP-TH/0410116;%%

%\cite{deHaro:2007eg}
\bibitem{deHaro:2007eg}
  S.~de Haro and P.~Gao,
  ``Electric - magnetic duality and deformations of three-dimensional CFT's,''
  arXiv:hep-th/0701144.
  %%CITATION = HEP-TH/0701144;%%

%\cite{Gaillard:1981rj}
\bibitem{Gaillard:1981rj}
  M.~K.~Gaillard and B.~Zumino,
  ``Duality Rotations For Interacting Fields,''
  Nucl.\ Phys.\  B {\bf 193}, 221 (1981).
  %%CITATION = NUPHA,B193,221;%%

%\cite{Gibbons:1995cv}
\bibitem{Gibbons:1995cv}
  G.~W.~Gibbons and D.~A.~Rasheed,
  ``Electric - magnetic duality rotations in nonlinear electrodynamics,''
  Nucl.\ Phys.\  B {\bf 454}, 185 (1995)
  [arXiv:hep-th/9506035].
  %%CITATION = NUPHA,B454,185;%%

%\cite{Tseytlin:1996it}
\bibitem{Tseytlin:1996it}
  A.~A.~Tseytlin,
  ``Self-duality of Born-Infeld action and Dirichlet 3-brane of type IIB
  superstring theory,''
  Nucl.\ Phys.\  B {\bf 469}, 51 (1996)
  [arXiv:hep-th/9602064].
  %%CITATION = NUPHA,B469,51;%%

%\cite{Gaillard:1997rt}
\bibitem{Gaillard:1997rt}
  M.~K.~Gaillard and B.~Zumino,
  ``Nonlinear electromagnetic self-duality and Legendre transformations,''
  arXiv:hep-th/9712103.
  %%CITATION = HEP-TH/9712103;%%

%\cite{Lutken:1991jk}
\bibitem{Lutken:1991jk}
  C.~A.~Lutken and G.~G.~Ross,
  ``Duality in the quantum Hall system,''
  Phys.\ Rev.\  B {\bf 45}, 11837 (1992)
  %%CITATION = OUTP-91-19-P;%%

%\cite{Lutken:1992xu}
\bibitem{Lutken:1992xu}
  C.~A.~Lutken and G.~G.~Ross,
  ``Delocalization, duality and scaling in the quantum Hall system,''
  Phys.\ Rev.\  B {\bf 48}, 2500 (1993)
  %%CITATION = OUTP-92-22-P;%%

%\cite{Burgess:2000kj}
\bibitem{Burgess:2000kj}
  C.~P.~Burgess and B.~P.~Dolan,
  ``Particle-vortex duality and the modular group: Applications to the  quantum
  Hall effect and other 2-D systems,''
  arXiv:hep-th/0010246.
  %%CITATION = HEP-TH/0010246;%%

%\cite{Burgess:2001sy}
\bibitem{Burgess:2001sy}
  C.~P.~Burgess and B.~P.~Dolan,
  ``Duality and Non-linear Response for Quantum Hall Systems,''
  Phys.\ Rev.\  B {\bf 65}, 155323 (2002)
  [arXiv:cond-mat/0105621].
  %%CITATION = PHRVA,B65,155323;%%

%\cite{Rozali:2007rx}
\bibitem{Rozali:2007rx}
  M.~Rozali, H.~H.~Shieh, M.~Van Raamsdonk and J.~Wu,
  ``Cold Nuclear Matter In Holographic QCD,''
  arXiv:0708.1322 [hep-th].
  %%CITATION = ARXIV:0708.1322;%%

%\cite{Henningson:1998gx}
\bibitem{Henningson:1998gx}
  M.~Henningson and K.~Skenderis,
  ``The holographic Weyl anomaly,''
  JHEP {\bf 9807}, 023 (1998)
  [arXiv:hep-th/9806087].
  %%CITATION = JHEPA,9807,023;%%

%\cite{Henningson:1998ey}
\bibitem{Henningson:1998ey}
  M.~Henningson and K.~Skenderis,
  ``Holography and the Weyl anomaly,''
  Fortsch.\ Phys.\  {\bf 48}, 125 (2000)
  [arXiv:hep-th/9812032].
  %%CITATION = FPYKA,48,125;%%

%\cite{Balasubramanian:1999re}
\bibitem{Balasubramanian:1999re}
  V.~Balasubramanian and P.~Kraus,
  ``A stress tensor for anti-de Sitter gravity,''
  Commun.\ Math.\ Phys.\  {\bf 208}, 413 (1999)
  [arXiv:hep-th/9902121].
  %%CITATION = CMPHA,208,413;%%

%\cite{deHaro:2000xn}
\bibitem{deHaro:2000xn}
  S.~de Haro, S.~N.~Solodukhin and K.~Skenderis,
  ``Holographic reconstruction of spacetime and renormalization in the  AdS/CFT
  correspondence,''
  Commun.\ Math.\ Phys.\  {\bf 217}, 595 (2001)
  [arXiv:hep-th/0002230].
  %%CITATION = CMPHA,217,595;%%

%\cite{Karch:2005ms}
\bibitem{Karch:2005ms}
  A.~Karch, A.~O'Bannon and K.~Skenderis,
  ``Holographic renormalization of probe D-branes in AdS/CFT,''
  JHEP {\bf 0604}, 015 (2006)
  [arXiv:hep-th/0512125].
  %%CITATION = JHEPA,0604,015;%%

\end{thebibliography}
\end{document}